\documentclass[prl,twocolumn,aps,superscriptaddress,showpacs,floatfix]{revtex4}
\usepackage{graphicx}
\usepackage{dcolumn}
\usepackage{bm}
\usepackage[usenames,dvipsnames]{color}
\usepackage{amsmath}
\begin{document}
\title{Stability of liquid films covered by a carpet of self-propelled surfactant particles}
%
\author{Andrey Pototsky}
\affiliation{Department of Mathematics, Faculty of Science Engineering and Technology, Swinburne University of Technology, Hawthorn, Victoria, 3122, Australia}
\author{Uwe Thiele}
\email{u.thiele@uni-muenster.de}
\homepage{http://www.uwethiele.de}
\affiliation{Institut f\"ur Theoretische Physik, Westf\"alische
 Wilhelms-Universit\"at M\"unster, Wilhelm Klemm Str.\ 9, D-48149 M\"unster, Germany}
\author{Holger Stark}
\email{Holger.Stark@tu-berlin.de}
\affiliation{Institut f{\"u}r Theoretische Physik, Technische Universit{\"a}t Berlin, Hardenbergstrasse 36, 10623, Berlin, Germany}
%
%
%
%
%

\begin{abstract}
We consider a carpet of self-propelled particles at the liquid-gas interface 
of a liquid film on a solid substrate. The particles excert an excess pressure
on the interface and also move along the interface while the swimming direction
performs rotational diffusion.
We study the intricate influence of these self-propelled insoluble surfactants on the stability of the film
surface and show that depending on the strength of in-surface rotational diffusion and the absolute 
value of the in-surface swimming velocity several characteristic instability modes can occur. In particular, 
rotational diffusion can either stabilize the film or induce instabilities of different character.
%
%

\end{abstract}

\maketitle
%
%
The understanding of the physical principles of the motion of
self-propelled particles in viscous fluids
\cite{Lauga09,Saintillan13,Aronson13,Kapral13, Zoettl14,Hennes14},
either in the bulk or at
interfaces is of primary importance for an increasing number of
applications in microfluidics and medicine
\cite{velev_nature07,Ayusman13,Ayusman13_b}.  A particularly
interesting emerging application of such swimmers are biocoatings
formed using a suspension of living cells that is deposited onto a solid
substrate before the solvent is removed, e.g., by evaporation. This
technique is used to fabricate bacterial carpets consisting of living
bacteria with rotating flagella that are attached head down to a
polymer layer \cite{Darnton04}.  The created homogeneous monolayer of
living cells is seen as a prototype of a novel biomaterial with
remarkable applications, e.g., as artificial skin, self-cleaning
coating, or biosensor
\cite{Efim09,Xu05,Bennett96,Weiss97,Fidaleo06}. Beside applications in
biotechnology, free liquid-gas interfaces loaded with motile bacteria
occur naturally,
for example, at the sea surface \cite{gladushev}.
Microswimmers at the interface of a thin liquid film also show interesting 
collective phenomena since even in a dilute suspension they interact with each other 
through the surface flow field initiated by gradients in the surface tension and curvatures in the height profile. 
In the following, we  present an analysis of the stability of such thin films and demonstrate
the subtle influence of in-plane swimming velocity and rotational diffusivity.

The proximity of swimmers to the liquid-gas interface inevitably
modifies the local surface tension, which depends on the swimmer
concentration similar to passive surfactant molecules and
(nano-)particles \cite{Rosen12,Bink2002cocis,ThAP2012pf,GaCS2012l}.
A gradient in the surface tension due to a non-uniform concentration generates
fluid flow at the surface (solutocapillary Marangoni effect), a phenomenon well
studied for passive surfactants.
%
The impact of self-propelled surfactants, i.e., surfactants that are capable to 
move autonomously, on the dynamics of liquid-gas interfaces (free surfaces), has been studied only in
one special case. Namely, Ref.~\cite{AlMi2009pre} investigates a monolayer of insoluble swimmers 
that are adsorbed at the free surface of a liquid film and exclusively swim into the direction perpendicular
to the surface.  That is, they are ``head up'', at all times and their motion along the liquid-gas interface is 
as for passive particles.  
As a result, the swimmers generate an excess normal pressure
  and fluctuations in their density may ``bulge" the interface
  locally. The induced fluid flow moves additional swimmers towards
  the bulge and increases the local excess pressure further. The
  interface becomes unstable if the combined stabilizing effect of
  translational diffusion and Marangoni flow towards regions of
  smaller swimmer concentrations is too weak\ \cite{AlMi2009pre}.

In this letter we show that self-propelled motion of the swimmer
parallel to the liquid-gas interface (in-plane motion) together with
rotational diffusion has a profound and non-trivial effect on the
stability of the film. 
Depending on the strength of rotational diffusion and swimming velocity, 
the in-plane motion can stabilize a flat film or 
induce film instabilities of different character.


Consider a liquid film on a smooth homogeneous solid substrate with
free liquid-gas interface. The liquid-gas interface is loaded with
self-propelled particles, each characterized by a unit vector ${\bm p}$ that gives the instantaneous 
direction of swimming with swimming velocity $v_0$.
\begin{figure}[ht]
\centering
\includegraphics[width=0.47\textwidth]{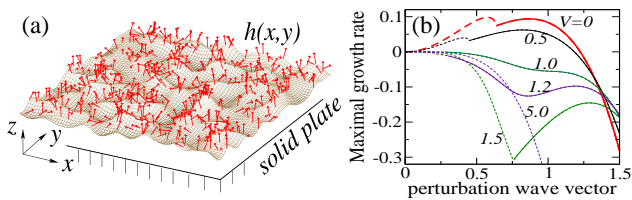}
\caption{(Color online) { (a) Liquid-gas interface of a liquid film, loaded
  with self-propelled surfactant particles. Their swimming orientations are indicated by the unit vectors ${\bm p}$, shown by red arrows. (b) 
  Influence of the in-line motion of self-propelled particles on film stability. 
  Largest growth rate of a perturbation around a flat film plotted versus wave number for different values of the in-plane 
  self-propulsion velocity $V$ as indicated. Other parameters are
  $D=1$ and $d=0$. Solid (dashed) lines correspond to oscillatory (steady) dynamics.}
  \label{F1}}
\end{figure}
%
%
For mean film thicknesses below several hundred micrometers, the
deformations of the liquid-gas interface are long-wave and can be
described 
in the long-wave or lubrication approximation
\cite{lubrication}. The position of a particle moving along the
liquid-gas interface is given by the three-dimensional (3d) vector 
$[{\bm  r}(t),h(x,y,t)]$, where ${\bm r}(t) = [x(t),y(t)]$  is a
two-dimensional (2d) position vector and $h(x,y,t)$ is the local film
height. In long-wave approximation, the projections of the orientation
vector ${\bm p}$ orthogonal  and parallel to the liquid-gas interface,
are approximated
by its $z$-component, $p_\perp \approx p_z $, and by ${\bm p}_\parallel \approx (p_x,p_y)$, respectively.  
Thus, the
overdamped equations of motion for the $i$-th particle become 
\begin{eqnarray}
\label{particles}
\dot{\bm r}_i&=&v_0({\bm p}_\parallel)_i+{\bm U}_i+{\bm \chi}_i(t),\nonumber \\
\dot{\bm p}_i&=&\left[{\bm \eta}_i + \frac{1}{2} {\bm \Omega}_i\right] \times {\bm p}_i, 
\end{eqnarray}
where ${\bm \chi}_i(t)$ with $\langle {\bm \chi}_i(t) \otimes {\bm
  \chi}_k(t^\prime)\rangle=2 M k_B T \delta_{ik}\delta(t-t^\prime){\bm 1}$
 and ${\bm \eta}_i$  with $\langle {\bm \eta}_i(t) \otimes {\bm \eta}_k(t^\prime)\rangle = 2D_r\delta_{ik}\delta(t-t^\prime){\bm 1}_3$
represent translational (2d) and rotational (3d) noise, respectively \cite{note_symbols}.
They both influence the dynamics of the self-propelled surfactants
at the liquid-gas interface. 
$M$ is the translational mobility and $D_r$ the rotational diffusivity
of the swimmers (see note~\cite{noteMobilities}),
${\bm U}_i = [U_x(x_i,y_i),U_y(x_i,y_i) ]$ is the surface velocity of the
fluid, and ${\bm \Omega}_i={\bm \nabla}\times {\bm U}_i=\Omega_i {\bm e}_z$
with $\Omega=\partial_x U_y -\partial_y U_x$ denotes the vorticity
at point $(x_i,y_i)$.  
%
For $\bm{\xi}_i =  {\bm \eta}_i = 0$, Eqs.\,(\ref{particles}) reduce to the case studied in
Ref.\,\cite{Zoettl12}, where a swimmer moves in a prescribed
time-independent Poiseuille 
flow field ${\bm U}$.
In our case, the flow field is initiated by variations in the liquid-film height and by the Marangoni effect.

So far, Eqs.\ (\ref{particles}) describe the dynamics of the full orientation vector ${\bm p}$ at the interface.
To reduce the dimensionality of the problem, we proceed to decouple
the dynamics of 
its in-plane component
${\bm p}_\parallel$ from the dynamics of the vertical orientation $p_\perp$.  In spherical coordinates, we have
$p_\perp \approx p_z = \cos{\theta}$ and ${\bm p}_\parallel \approx
(p_x,p_y)=\sin{\theta} \,\mathbf{q}$ with ${\bm q}=(\cos{\phi},\sin{\phi})$.
%
Assuming that the characteristic rotational diffusion time of the
swimmers $\tau\sim D_r^{-1}$ is much smaller than the characteristic
relaxation time of the film thickness fluctuations (a plausible
assumption for the quasi-stationary lubrication approximation), one
may safely assume that the distribution of the vertical orientation of
swimmers $p_z$ adjusts ``instantaneously'' to some stationary
distribution $P(p_z)$. This allows us to average
Eqs.\,(\ref{particles}) over the angle $\theta$ to obtain
%
%
%
\begin{eqnarray}
\label{particles_2}
\dot{\bm r}_i&=&v\mathbf{q}_i+{\bm U}_i +{\bm \chi}_i(t),\nonumber \\
\dot{\phi}_i&=&\frac{1}{2} \mid {\bm \Omega}_i \mid + \xi_i(t),
\end{eqnarray}
where $v=v_0 \langle \sqrt{1-p_z^2}\rangle$, with $\langle \sqrt{1-p_z^2}\rangle = \int \sqrt{1-p_z^2}\,P(p_z)\,dp_z$ representing the average 
absolute value of ${\bm p}_{\parallel}$, and $\xi(t)$ is a Gaussian white noise with $\langle \xi_i(t)\xi_k(t^\prime)\rangle = 2D_r \delta_{ik}\delta(t-t^\prime)$. 
%
%
Note that Eqs.\,(\ref{particles_2}) describe the 2d 
dynamics of a self-propelled particle with propulsion speed $v$ moving in a fluid with
velocity field ${\bm U}=(U_x,U_y)$. 

The Smoluchowski equation for the surface particle density $\rho({\bm  r},\phi,t)$ corresponding to Eqs.~(\ref{particles_2}) is 
a continuity equation with translational and orientational currents that, as usual, contain drift and diffusional contributions:
%
\begin{equation}
\label{Smoluchowski}
\frac{\partial \rho}{\partial t} + {\bm \nabla}\cdot\left(v\rho{\bm q}+{\bm U}\rho \right)+\frac{\Omega}{2}\frac{\partial \rho }{\partial \phi} - D_r \frac{\partial^2 \rho}{\partial \phi^2}-{ M k_B T\Delta \rho}=0.
\end{equation}
Here, ${\bm \nabla}$ and $\Delta$ denote, respectively, the nabla and Laplace operator in positional coordinates.
Eq.\,(\ref{Smoluchowski}) is coupled to the thin film equation via the 
solutocapillary Marangoni effect.
Following Refs.\,\cite{AlMi2009pre,lubrication}, the equation for the local film thickness $h(x,y,t)$ is given by
\begin{equation}
\label{thin_film}
\frac{\partial h}{\partial t}+{\bm \nabla}\cdot\left( \frac{h^3}{3\mu}{\bm \nabla}\left[\Sigma_0 \Delta h + \alpha \langle \rho \rangle \right]\right) + {\bm \nabla}\cdot\left(\frac{h^2}{2\mu}{\bm \nabla}\Sigma\right)=0,
\end{equation}
where $\mu$ is the dynamic viscosity, $\Sigma(x,y,t)$ is the concentration-dependent surface
tension, $\Sigma_0$ is the constant reference surface tension without surfactants
(see note \cite{noteVariational}),  and $\alpha \langle \rho \rangle$ describes the additional pressure exerted by the
swimming particles onto the liquid-gas interface as in Ref.~\cite{AlMi2009pre}. Here, 
$\langle\rho\rangle(x,y,t)=\int_{0}^{2\pi} \rho(x,y,\phi,t)\,d\phi$ 
denotes the local particle density averaged over all swimming directions and 
$\alpha=\langle p_z \rangle v_0 /M $ with $\langle p_z \rangle = \int
P(p_z) p_z\,dp_z$ is the negative of the force the interface exerts on the swimming particles in order to stop them in vertical direction
 \cite{note_alpha}.
Finally, the fluid velocity field at the free surface reads \cite{lubrication}
\begin{equation}
\label{surface_vel}
{\bm U}=\frac{h}{\mu}{\bm \nabla}\Sigma + \frac{h^2}{2\mu}{\bm \nabla}\left(\Sigma_0 \Delta h + \alpha\langle \rho \rangle \right).
\end{equation}

%



To close our set of equations, we link surface tension $\Sigma$ to particle density $\rho$.
As any passive surfactant, self-propelled particles at the interface modify
the local surface tension $\Sigma$. Here, we assume that the concentration is low, i.e., the
surfactant particles are in a 2d gaseous state implying that $\Sigma$
depends linearly on the local particle concentration $\langle \rho
\rangle$ (cf.~Ref.~\cite{ThAP2012pf}):
\begin{equation}
\label{Marangoni}
\Sigma = \Sigma_0 -\Gamma \langle \rho \rangle.
\end{equation}
%
Typically, $\Gamma>0$, i.e., the surface tension decreases with increasing $\langle \rho \rangle$ 
and the Marangoni flow is directed towards lower particle concentration.



Equations~(\ref{Smoluchowski}) to~(\ref{Marangoni}) form a closed system
of nonlinear integro-differential equations for the two scalar fields $h(x,y,t)$ and $\rho(x,y,\phi,t)$. 
It is important to note that for $v=0$ and $\alpha=0$, the density is
independent of the angle $\phi$, i.e., $\rho=\rho(x,y,t)$ and
Eqs.~(\ref{Smoluchowski}) to 
(\ref{Marangoni}) 
reduce to the two usual coupled equations for film height $h(x,y,t)$ and surfactant
concentration $\rho(x,y,t)$ for passive insoluble surfactant
\cite{Schwartz95,lubrication}.

To non-dimensionalize, we use $h_0$ and $h_0\sqrt{\Sigma_0/\Gamma\rho_0}$ as the vertical 
and horizontal length scale, respectively, $\mu
h_0 \Sigma_0/(\Gamma^2 \rho_0^2)$ as the time scale and the
direction-averaged density of swimmers in the homogeneous state
$\langle \rho\rangle_0=\int_{0}^{2\pi}\rho_0/(2\pi)\,d\phi=\rho_0$ as
the density scale. We introduce the dimensionless in-plane
self-propulsion velocity $V=v\mu\Sigma_0^{1/2}/(\Gamma\rho_0)^{3/2}$,
rotational diffusivity $D=D_r h_0 \mu\Sigma_0/(\Gamma\rho_0)^2$
\cite{note_diffusion},
surface diffusivity $d=k_B T M \mu/(h_0\rho_0\Gamma)$, and excess pressure  $\beta=\alpha h_0 /\Gamma$.
The dimensionless equations of motion derived from
Eqs.\,(\ref{Smoluchowski}), (\ref{thin_film}), and (\ref{surface_vel})
are summarized in the Supplement.
%
%

The trivial homogeneous stationary state corresponds to a flat film
covered  uniformly by particles, i.e., $\rho=1/(2\pi)$ and $h=1$.
Any small amplitude perturbation of the trivial uniform state can be represented as
%
%
\begin{eqnarray}
\label{Fourier}
\delta
h({\bm r},t)&=& \int \hat{h}({\bm k})e^{\gamma({\bm k})t}e^{I{\bm k}{\bm r}}\,d{\bm k},\\
\delta
\rho({\bm r},\phi,t)&=&\lim_{N\rightarrow \infty} \frac{1}{2\pi}\sum_{n=-N}^{N} e^{I n \phi}\int W_n({\bm k})e^{\gamma({\bm k})t}e^{I{\bm k}{\bm r}}\,d{\bm k},
\nonumber
\end{eqnarray}
with small amplitudes $\hat{h}({\bm k})$ and $W_n({\bm k})$, the wave vector ${\bm k}=(k_x,k_y)$, and the growth rate $\gamma({\bm k})$.
For any fixed number $N$ of the Fourier modes, the growth rate $\gamma({\bm k})$ of the fastest growing perturbation is found 
  by solving the 
  eigenvalue problem obtained by linearizing the dimensionless Eqs.\,(\ref{Smoluchowski}), (\ref{thin_film}), and
  (\ref{surface_vel}), as outlined in the Supplement. For the
  parameter values used here, the results have converged for $N=10$.

\begin{figure}[ht]
\centering
\includegraphics[width=0.49\textwidth]{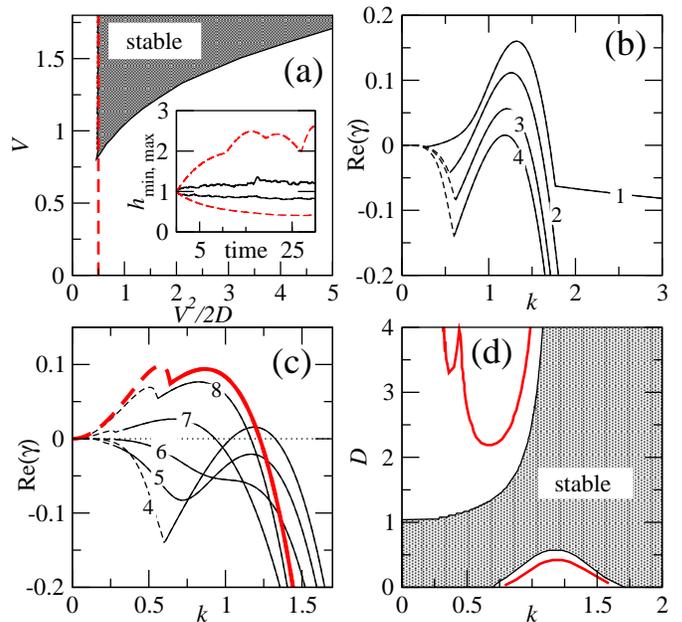}
\caption{(Color online) 
(a) Stability diagram in the plane $(d_{\mathrm{eff}}=V^2 /
  2D,V)$ at fixed $\beta=3$ and $d=0$. The inset shows the time evolution of the minimum 
 and maximum of the local film thickness in the Smooth Particle Dynamics simulations for 
 $V=0$ (dashed lines) and $V=5$ (solid lines) at $D=1$.  
(b) and (c) The dispersion relations (maximal $Re(\gamma(k))$) for $V=1$ and  different
$D$ as indicated by numbers near each curve: 1: $D =
0.001$, 2: $D = 0.1$, 3: $D = 0.3$, 4: $D = 0.5$, 
5: $D = 0.7$, 6: $D = 1$, 7:$D = 2$, and 8: $D = 10$. Solid (dashed) lines correspond to oscillatory (steady) perturbations. The heavy solid
line represents $D \to \infty$ and coincides with the
curve for $V = 0$ in Fig.~\ref{F1}(b). 
(d) Stability diagram in the plane $(D,k)$. The solid line represents the level line 
of the dispersion relation at Re$(\gamma)=0.03$ to illustrate its bimodal character.
\label{F2}}
\end{figure}
%


In what follows, we set $d=0$ and focus on the most striking effects
of the in-plane motion and in-plane rotational diffusion of the
swimmers on film stability. The choice of $d=0$ is motivated by
estimating the ratio 
$V^2/(D d) = v^2/ (D_r M k_BT) \gg 1$ using thermal values for both diffusion coefficients
with $R = 1 \mathrm{\mu m}$ and $v = 1 \mathrm{\mu m/s}$.
As the eigenvalues $\gamma({\bm k})$ do only depend on the absolute value of ${\bm k}$ 
but not on its direction (spatial isotropy), we compute the growth rate $Re(\gamma)$ as 
a function of $k$.


As a reference we set the excess pressure parameter $\beta=3$
to be slightly above the critical value of $\beta_c=2$, where the
long-wave steady instability sets in for $V=0$. This case
corresponds to particles that only swim perpendicularly to the free
surface as studied in Ref.~\cite{AlMi2009pre}.  The corresponding
dispersion relation $Re(\gamma)$ \textit{vs.}\ $k$ is plotted as heavy
solid line in Fig.~\ref{F1}(b) and indicates that the excess pressure
due to the upwards swimming surfactants drives the film unstable.
Fig.~\ref{F1}(b) also shows how the film is stabilized by the 
in-plane self-propelled particle motion, i.e., when increasing $V$ from zero
(here, at fixed rotational diffusion $D=1$).
We understand this stabilization qualitatively since active motion along the interface
acts like the stabilizing translational diffusion with an effective diffusivity $d_{\mathrm{eff}}=V^2 / 2D$
on time scales larger than the orientational correlation time $D^{-1}$  \cite{Howse07,Palacci10,EncStark11}.
We further note that at low $V$ the dispersion relation has two maxima
that correspond to a
steady (growing drops and holes) and an oscillatory instability (traveling waves) mode that dominate at $V=0$ and $V=0.5$,
respectively. The onset of the steady instability occurs at $V\approx1$ at
zero wave number.


%

The accompanying stability diagram in the ($V, d_{\mathrm{eff}}$)-parameter plane [Fig.~\ref{F2}(a)] provides
  quantitative insight into the stabilization of the liquid film. It
  reveals an intermediate range of $d_{\mathrm{eff}}=V^2 / 2D$ where
  the film is stable. A small effective diffusivity
  $d_{\mathrm{eff}}$ (large $D$) cannot stabilize the
  film. The film becomes stable when $d_{\mathrm{eff}}$ exceeds a
  threshold value, which exactly agrees with the stabilizing
  translational diffusivity obtained in Ref.~\cite{AlMi2009pre}
  for swimmers with purely head-up orientation (red dashed line).
  Indeed, by expanding the full density $\rho(x,y,\phi,t)$ into
  angular moments and deriving dynamic equations for the moments from
  Eq.\ (\ref{Smoluchowski}), one can formulate a
  Smoluchowski equation for $\langle\rho\rangle(x,y,t)=\int_{0}^{2\pi}
  \rho(x,y,\phi,t)\,d\phi$ on time scales larger than $D^{-1}$ and on
  large length scales, where the active motion only contributes to an
  effective translational diffusion constant $d+d_{\mathrm{eff}}$
  \cite{Golestanian12,Pohl14}. So, one obtains the density equation
  employed in Ref.\ \cite{Schwartz95} for passive surfactants and in
  Ref.\ \cite{AlMi2009pre} for swimmers with purely head-up
  orientation.

The selected dispersion relations for the growth rate in
Fig.~\ref{F2}(b) and (c) at $V=1$ reveal a small wave length
instability for $D \approx 0$ (curve 1) to $D \approx 0.6$, which
corresponds to the unstable region in (a) for large
  $d_{\mathrm{eff}}$. A stability analysis of our dynamic equations
  for $D=0$ at a fixed delta-peaked distribution of the swimming
  direction reveals a growing density and height modulation wave that
  travels with the swimming speed $V$. This implies that dense swimmer
  regions do not disperse and render the film unstable. With
  increasing $D$ the swimming direction starts to diffuse what
  suppresses the traveling waves more and more (depending on $k$)
  until at $D \approx 0.6$ the trivial state is stable. The stability
  diagram in the $(k,D)$-plane in Fig.~\ref{F2}(d) illustrates the
  qualitatively different character of the instabilities at low and
  large $D$. Above the stable range at intermediate $D$, the film
  becomes again unstable when the effective translational diffusivity
  $V^2/2D$ becomes too small. The onset occurs at $k=0$ as a steady
  instability, however, with further increasing $D$ the dispersion
  relation develops a bimodal character as indicated in
  Fig.~\ref{F2}(d) by the shown level line. Note, finally, that at
  large $D$ [$V^2/D \rightarrow 0$], the dispersion relation is
  identical to the one in the unstable reference case of purely
  perpendicular swimming [red curve in Fig.~\ref{F2}(c)].

In order to confirm the predicted stabilization of a liquid film by
motile surfactant particles, we numerically solve the thin film
equation [Eq.\,(\ref{thin_film})] in a $L\times L$ square box with
periodic boundary conditions, coupled to the equations of motion
[Eqs.\,(\ref{particles_2})] for $n=500$ individual point swimmers.
Their discrete spatial distribution is translated into a smooth
particle density function $\rho(x,y,t)$ employing the method of the
Smooth Particle Dynamics \cite{Hoover} (for details see the
Supplement).
%
%

Fixing the remaining parameters as in Fig.\ref{F1}(b), we vary
the self-propulsion velocity $V$. From the linear stability analysis
we expect that the flat film is linearly unstable for $V=0$ and linearly stable for $V=5$. 
Starting from randomly located and oriented particles on the
  flat film surface, we observe a time evolution that is noisy in
both cases due to the coupling of the continuum equation to the
discrete particle dynamics. This can be appreciated in the inset of
Fig.~\ref{F2}(a) showing the time evolution of the minimum and maximum
of the local film thickness for $V=0$ (dashed lines) and $V=5$ (solid
lines). Corresponding movies are available in the Supplement.

One discerns a clear difference between the two cases: For
  $V=0$, the swimmer-induced instability of the flat film is
  apparent. The amplitude of the surface deflections first grows
  before at later times it varies (in a potentially chaotic way) about
  a maximum that is by about a factor 2 larger than the mean film
  thickness $h=1$. This resembles a regime of interacting nonlinear
  'traveling waves' as predicted in Ref.~\cite{AlMi2009pre}.
  However, for swimmers with in-plane motility (here, $V=5$), one
  finds that after a small initial growth the deformation amplitude
  soon saturates and then fluctuates about a (small) finite amplitude
  that is by about a factor 2 smaller than $h=1$. This indicates that
  the instability is strongly suppressed. The remaining fluctuations
  of the film surface are a consequence of the hybrid
  calculations where the discrete stochastic dynamics of the finite
  number of swimmers is coupled to the continuum model for the
  evolution of the film height.

In conclusion, we have shown that rotational diffusion can play
  a distinctive role in the motion of self-propelled surfactants on
  liquid films. Depending on its relative strength, it can stabilize or
  destabilize the film and may also change the qualitative character
  of the instability and, in consequence, the nonlinear behaviour.
  More specifically, our analysis shows the flat film can be
  destabilized 
 according to two different scenario: (i)
  through the steady long wave instability at $V^2/2D=0.5$ and (ii)
  through the oscillatory finite wave length instability along the
  solid line in Fig.\,\ref{F2}(a).

However, one needs to show that such a system is experimentally
  feasible as was done for the ``head-up" self-propelled particles
  in Ref.\,\cite{AlMi2009pre}. Next we assess whether this also
  applies for the stabilization of the film due to the combined
  action of the rotational diffusivity $D$ and the in-plane velocity
  $V$, reported here. As an example, we estimate the dimensionless $V$ and $D$ for
  self-propelled Janus particles \cite{bechinger12} in a water film at
  room temperature: thus we take $\mu=10^{-3}$ kg m$^{-1}$s$^{-1}$,
  $\Sigma_0=10^{-1}$ N m$^{-1}$, $\Gamma=10^{3}$ kg m$^2$s$^{-2}$mol$^{-1}$
  (from \cite{AlMi2009pre}), $v=10^{-6}$ ms$^{-1}$, $R=10^{-6}$
  m (from \cite{bechinger12}). In addition, we estimate the (largest
  possible) average density $\rho_0\approx(\pi R^2)^{-1}\approx
  10^{-12}$ mol m$^{-2}$ and take $h_0=10^{-4}$ m.  Taking into account
  that for a spherical Brownian particle $D_r=3k_BT M/4R^2$ and
  $M=(6\pi \mu R)^{-1}$ we obtain $V\sim 10^{4}$ and $V^2/D \sim
  10^{-1}$. These estimates show that 
self-propelled Janus particles
  in a $100$ $\mu$m thick water film approximately fall onto the
  vertical dashed line in Fig.\,\ref{F2}(a). Consequently, by fine
  tuning the radius $R$ of the particles ($V \propto R^3$ and $V^2/D \propto R^5$)
  or the self-propulsion velocity $v$, one can induce or suppress the long-wave steady
  instability of the film.


\acknowledgments
A.P. thanks the research training group GRK 1558 funded by DFG
for financial support.


\begin{thebibliography}{37}
\bibitem[{not({\natexlab{a}})}]{note_symbols}
\bibinfo{note}{Here $\otimes$ stands for the tensor product between two
  vectors, ${\bm 1}$ is the unit $2\times 2$ matrix and ${\bm 1}_3$ is the unit
  $3\times 3$ matrix}.
\bibitem[{not({\natexlab{d}})}]{note_alpha}
\bibinfo{note}{Note that $\alpha$ depends on the mobility of the particle $M$,
  which in its turn is linked to the the particle radius via the Stokes' law,
  i.e. $M=(6\pi\mu R)^{-1}$ for spherical particles of radius $R$.}
\bibitem[{not({\natexlab{e}})}]{note_diffusion}
\bibinfo{note}{Note that for the dimensionless thermal rotational diffusion
  $D = k_BT\sigma_0 / (8\pi\Gamma^2\rho_0^2) \times h_0 / R^3$,
  where $R$ is the radius of the spherical swimmer.}
\end{thebibliography}
\end{document}